\documentclass[final]{svjour2}
\usepackage{graphicx}
\usepackage{rotating}
\usepackage{amssymb}
\usepackage{mathptmx}
\usepackage[numbers]{natbib}
\makeatletter
\journalname{Journal of Low Temperature Physics}
\begin{document}

\newcommand{\hdblarrow}{H\makebox[0.9ex][l]{$\downdownarrows$}-}
\title{Low-noise HEMTs for Coherent Elastic Neutrino Scattering and Low-Mass Dark Matter Cryogenic Semiconductor Detectors}

\author{A. Juillard, J. Billard, D. Chaize, J-B Filippini, D. Misiak, L. Vagneron, A. Cavanna, Q. Dong, Y. Jin, C. Ulysse, A. Bounab, X. de la Broise, C. Nones, A. Phipps}

\institute{A. Juillard, J. Billard, D. Chaize, J-B Filippini, D. Misiak, L. Vagneron \at Univ. Lyon, Univ. Lyon 1, CNRS/IN2P3, IPN-Lyon, F-69622, Villeurbanne, France  %\\ Tel.:\
\email{alexandre.juillard@ipnl.in2p3.fr}
\and
Q. Dong, A. Cavanna, C. Ulysse, Y. Jin \at C2N, CNRS, Univ. Paris-Sud, Univ. Paris-Saclay, 91120 Palaiseau, France
\and
A. Bounab, X. de la Broise, C. Nones \at
IRFU, CEA, Univ Paris-Saclay , F-91191 Gif-sur-Yvette, France
\and
A. Phipps \at
Department of Physics, Stanford University, Stanford, CA 94305, USA
}
\maketitle

\maketitle

\begin{abstract}

We present the noise performance of High Electron Mobility Transistors (HEMT) developed by CNRS/C2N laboratory. Various HEMT's gate geometries with 2 pF to 230 pF input capacitance have been studied at 4 K. A model for both voltage and current noises has been developed with frequency dependence up to 1 MHz. These HEMTs exhibit low dissipation, excellent noise performance and can advantageously replace traditional Si-JFETs for the readout of high impedance thermal sensor and semiconductor ionization cryogenic detectors.
Our model predicts that cryogenic germanium detectors of 30 g with 10 eV heat and 20 eV$_{ee}$ baseline resolution are feasible if read out by HEMT-based amplifiers.
Such resolution allows for high discrimination between nuclear and electron recoils at low threshold. This capability is of major interest for Coherent Elastic Neutrino Scattering and low-mass dark matter experiments such as Ricochet and EDELWEISS.


\end{abstract}

\section{Introduction}

High-purity cryogenic semiconductor detectors with simultaneous readout of phonon and ionization signals have been developed over the past three decades. A nuclear recoil produces three to ten times less ionization in germanium or silicon than an electron recoil does. This allows for an excellent event-by-event discrimination between nuclear and electron recoils.
Discrimination power better than $10^6$ have been demonstrated by both the CDMS and EDELWEISS \cite{EDW_JINST} collaborations for recoil energies above $\sim$ 10 keV.

These developments were initially motivated by the direct search for Dark Matter in the form of Weakly Interacting Massive Particles (WIMPs). In the 1-100 GeV range the detection consists of measuring the recoil energy produced by the elastic scattering of a WIMP off target nuclei. The signal is thus a nuclear recoil while the background are electron recoils induced by $\alpha$, $\beta$ and $\gamma$ radioactivity. 
Due to the low ionization yield for nuclear recoils the detection threshold is driven by the heat channel while the discrimination threshold is limited by the ionization channel resolution. 

The ionization baseline resolution is limited by the noise of the Si-JFET transistor traditionally used for the input stage of charge amplifier. The commonly achieved baseline resolution on massive cryogenic detectors is $\sigma_E\approx.3\,$ keV$_{ee}$ (keV-electron equivalent, ionization energy produced by a 1 keV electronic recoil). This corresponds to the maximum ionization signal of a 3 GeV WIMP, taking into account the kinematic and Dark Matter halo properties . At lower masses the discrimination power of cryogenic semiconductor detectors vanishes and electron recoil background becomes the main limitation to the sensitivity.

First probed by 10 kg-scale cryogenic detector experiments \cite{EDW_HM, EDW_LM, CDMS_HM, CDMS_LM, CRESST_HM, CRESST_LM} the 5-100 GeV WIMP mass range region is now extensively explored by ton-scale liquid noble gas detectors such as LUX \cite{LUX}, PandaX \cite{PandaX} and XENON \cite{XENON1T} with no detection yet. This has raised an increasing interest in low-mass ($<$ 5 GeV) WIMP searches  \cite{CRESST_LM, EDW_Surf, EDW_proj, CDMS_proj, NEWSG, DAMIC} and motivated novel techniques to lower thresholds including single electron-hole sensitivity \cite{CDMSsingle, SkipperCCD}, however at the price of no more or very poor background discrimination.

Low thresholds also enable cryogenic semiconductor detectors to study Coherent Elastic Neutrino-Nucleus Scattering (CE$\nu$NS). The COHERENT collaboration recently reported the first observation of CEvNS from spallation-sourced ~30 MeV neutrinos \cite{Coherent}. In germanium, the recoil energy spectrum for lower-energy 5 MeV neutrinos is similar to that of a 3.5 GeV WIMP. The cryogenic Dark Matter detector community has therefore started to work on three projects, Richochet, NUCLEUS and MINER \cite{Ricochet, Nucleus, Miner}, to measure CE$\nu$NS from few MeV neutrinos produced by nuclear reactors. Beyond confirming the first observation of CE$\nu$NS, the low energy threshold of these experiments will be uniquely position them to test a myriad of exotic physics scenarios.

In this paper, we show that High Electron Mobility Transistors (HEMTs) developed by the CNRS/C2N laboratory enable the development of low threshold, high discrimination germanium detectors for the EDELWEISS-SubGeV Dark Matter program and Ricochet CE$\nu$NS experiment. The superior operational performance of the HEMTs allows the detectors to measure sub-keV nuclear recoils induced by sub-GeV WIMPs and MeV neutrinos while retaining electromagnetic background discrimination.

\section{HEMT developed by CNRS/C2N}

\begin{figure} %[htbp]

\begin{center}
\includegraphics[%
  width=0.89\linewidth,
  keepaspectratio]{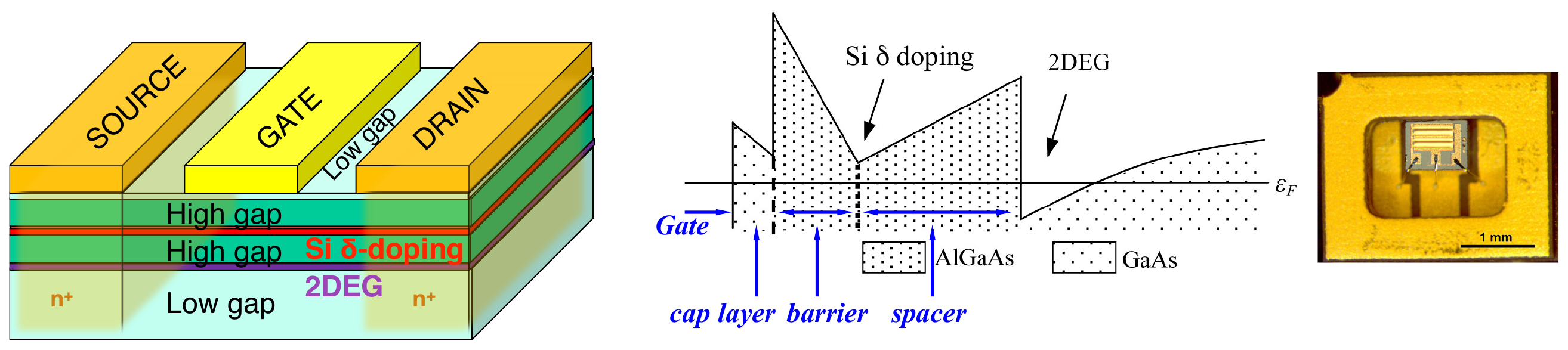}
\end{center}
\caption{{\it Left:} Schematic view of AlGaAs/GaAs heterostructure forming the HEMT. The high band gap, low band gap and Si-$\delta$ doping layers are shown. {\it Middle:} Associated energy band diagram seen by the gate. {\it Right:} 32 $\mu$m $\times$2 mm device in its SOT23 ceramic package. Wirebounds connecting the drain, source and gate contacts to the package are visible. (Color figure online.)}
\label{HEMTbasics}
\end{figure}

HEMTs are based on a 2-Dimensional Electron Gas (2DEG) realized in a heterostructure with a high purity material interface. The HEMTs produced for this work are based on AlGaAs/GaAs high gap/low gap heterostructures grown by Molecular Beam Epitaxy (MBE). The structure consists of a GaAs buffer layer, a 20 nm AlGaAs spacer layer (thicker than the few nm employed for commercial HEMTs), a Si $\delta$-doping layer, a 15 nm undoped AlGaAs barrier layer, and finally a 6 nm undoped GaAs cap layer. 
The high gap layers are fully depleted and donate electrons to the low gap buffer layer, forming a 2DEG at the heterostructure interface. Drain and source connections to the 2DEG are formed by ohmic contact through diffusion of Ge under an Au layer. The gate connection is a Schottky contact formed by an Au surface layer with a Ti underlayer to avoid diffusion.
When a negative bias is applied to the gate contact, electrons from the 2DEG are extracted to compensate the change in the gate charge. The drain-source current formed by the 2DEG can thus be modulated by the gate voltage: HEMTs are field effect transistors. A schematic view of the heterostructure and the associated energy band diagram seen by the gate are shown in Fig.~\ref{HEMTbasics}.

Unlike Si JFETs, whose conducting channel freezes out at low temperature, the 2DEG of the HEMT is formed even at T=0 K. As the HEMTs do not require a heating stage and dissipate significantly less power, they can be placed much closer to the detector. Closer proximity to the detector reduces the two dominant performance limitations of amplifiers for high impedance sensors, stray capacitance and microphonics susceptibility. 

HEMTs with various gate lengths and gate widths have been fabricated and individually packaged in a ceramic SOT23 package.
Five HEMT geometries with gate surfaces of 0.15 mm$^2$, 0.064 mm$^2$, 0.02 mm$^2$, 2$\times$10$^{-3}$ mm$^2$ and 4$\times$10$^{-4}$ mm$^2$ have been studied with input capacitance C$_{gs}$ of 230 pF, 100 pF, 30 pF, 5 pF and 2 pF, respectively.

\begin{figure} %[htbp]
\begin{center}
\includegraphics[%
  width=0.8\linewidth,
  keepaspectratio]{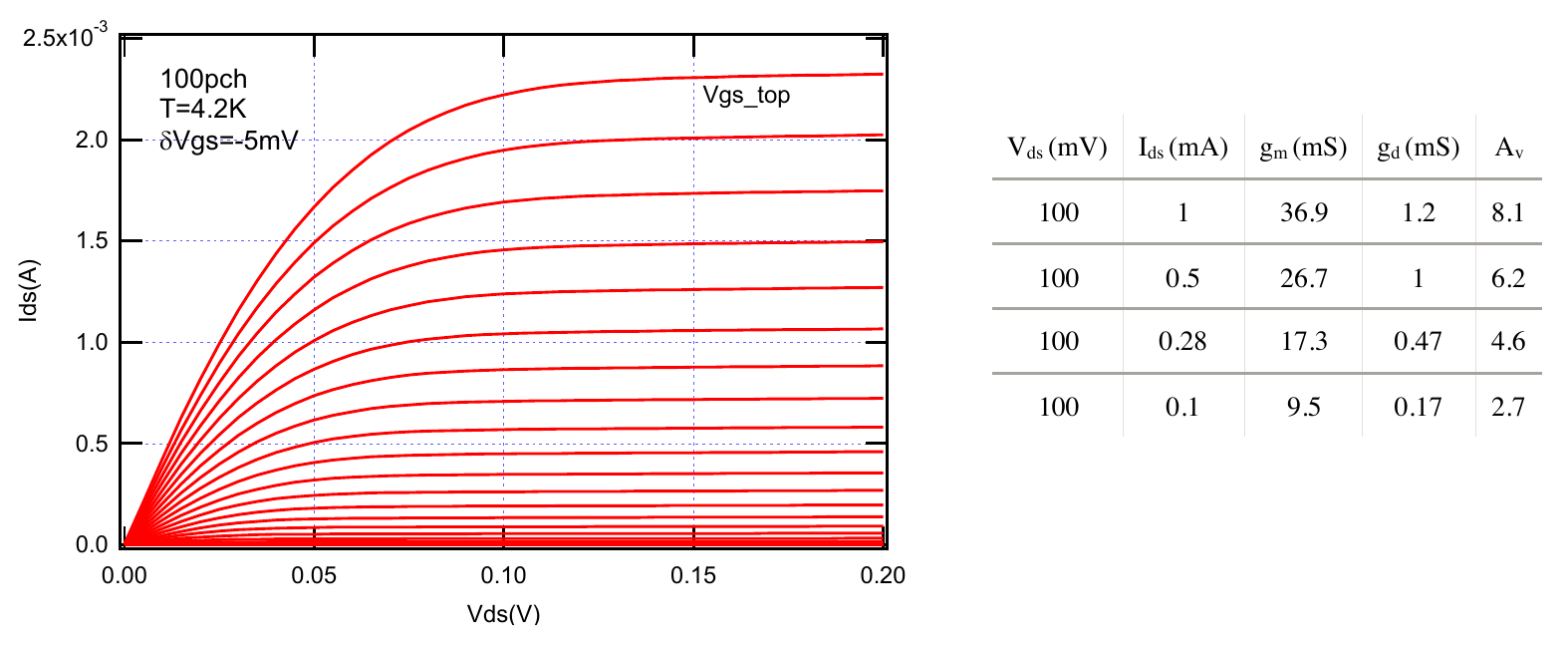}
\end{center}
\caption{{\it Left:} I$_{ds}$ vs V$_{ds}$ characteristics at 4.2K for a 100 pF HEMT. The top curve is measured with V$_{gs\_top}$ = -80 mV and from top to bottom a $\delta$V$_{gs}$ = -5 mV step is applied between each curve. {\it Right:} Transconductance, output conductance and voltage gain (measured in standard common source amplifier configuration with 300 $\Omega$ load resistance on the drain) at various working point. (Color figure online.)}
\label{IdeV}
\end{figure}

As an example, typical I$_{ds}$ vs V$_{ds}$ characteristics for various V$_{gs}$ are given in Fig.~\ref{IdeV} for C$_{gs}$ = 100 pF device at 4.2 K. V$_{gs\_top}$ can vary slightly from device to device and is typically around -80 mV. Transconductance g$_{m}$, output conductance g$_{d}$ and voltage gain A$_{V}$ have been studied at various saturation current.
Voltage gain A$_{V}$ is measured in standard common source amplifier configuration with 300 $\Omega$ load resistance on the drain \cite{HEMT_YongAPL}. 
High transconductance and thus reasonable voltage gain can be obtained for dissipation powers in the 10-100 $\mu$W range which is much lower than the few mW needed for Si-JFETs.

\section{HEMT noise : T = 4.2 K Measurements and Model}

\begin{figure} %[htbp]
\begin{center}
\includegraphics[%
  width=0.7\linewidth,
  keepaspectratio]{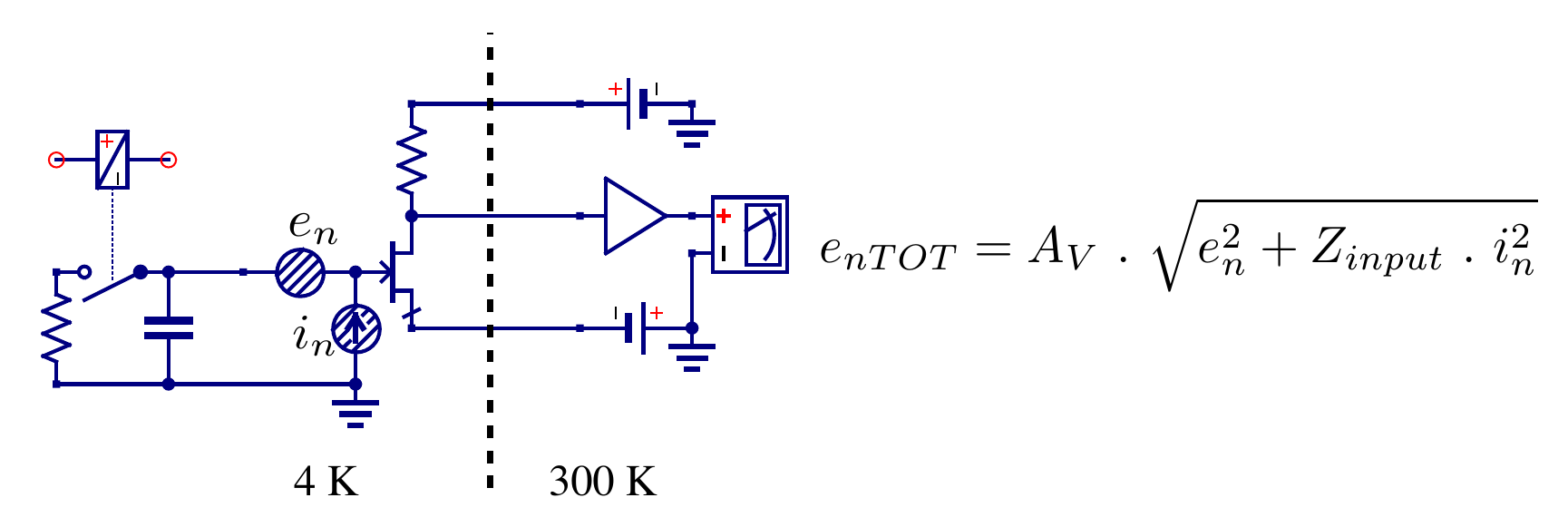}
\end{center}
\caption{Experimental setup used for voltage and noise measurement. Input impedance can be switched from low value resistance to pure capacitance thanks to a mechanical relay. Total noise and voltage gain are measured in common source amplifier configuration. (Color figure online.)}
\label{ComSourceAmp}
\end{figure}

The noise measurement procedure has been explained in details in previous work \cite{HEMT_YongAPL, HEMT_ICSICT} and is summarized in Fig.~\ref{ComSourceAmp}. A cold mechanical relay allows the amplifer load to be switched between a low resistance and various capacitors. Intrinsic voltage noise is measured with the low resistance input while current noise is extracted with total noise measurement through $e_{n TOT}= A_V~.~\sqrt{e_{n}^{2}+Z_{input}~.~i_{n}^{2}}$. Measurements of the noise power spectrum between 1 Hz and 1 MHz are referenced to the HEMT input voltage after division by the measured voltage gain $A_V$. 

\begin{figure} %[htbp]
\begin{center}
\includegraphics[%
  width=0.9\linewidth,
  keepaspectratio]{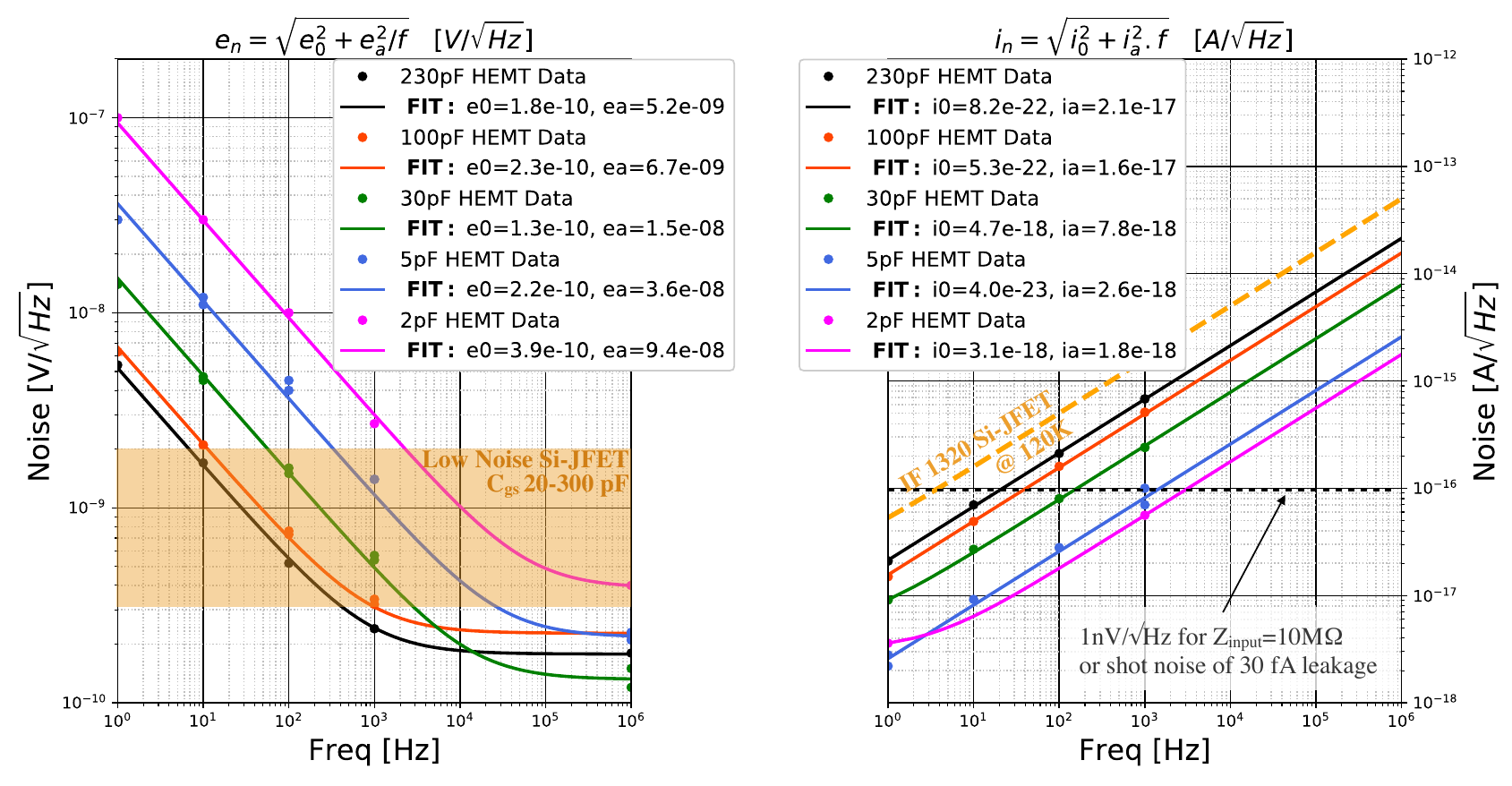}
\end{center}
\caption{{\it Left:} Voltage noise of the 5 HEMT geometries studied at 4.2 K and 100$\mu$W. Brown area illustrates the best voltage noise achieved by Si-JFET. {\it Right:} Current noise of the 5 HEMT geometries studied at 4.2 K and 100$\mu$W. Current noise of the IF1320 from InterFET studied at 120 K (data from \cite{EDWreadout}) is also plotted for comparison. Black dashed line at 10$^{-16}$/$\sqrt{Hz}$ is equivalent to the shot noise induced by a 30 fA leakage current and would gives a 1 nV/$\sqrt{Hz}$ voltage noise for a 10 M$\Omega$ input impedance. (Color figure online.)}
\label{EnIn}
\end{figure}

Results for both voltage and current noise for the five gate geometries are given in Fig.~\ref{EnIn} with the same 100 $\mu$W dissipation ( I$_{ds}$ = 1mA, V$_{ds}$ = 100mV) at 4.2 K. Voltage noises are compared to the very best commercially available Si-JFET from InterFET \cite{InterFET} which can have noise as low as 0.3-2 nV/$\sqrt{Hz}$ down to 1 Hz depending on their input capacitance.
As seen in Fig. 4, HEMTs from CNRS/C2N exhibit better or similar white noise performance to comparable Si-JFETs from Interfet, although the 1/f knee occurs at higher frequency. For all gate geometries, the voltage noise is well-fit by a 1/$\sqrt{f}$ term and a white noise term as in previous work \cite{HEMT_YongAPL, BerkeleyLTD16}. The noise magnitude varies as $\sqrt{C_{gs}}$.

Typical HEMTs suffer from poor 1/f noise up to several MHz. In this regard, the voltage noise performance of the CNRS/C2N HEMTs is greatly improved. For high impedance sensors (i.e. ionization), however, the main advantage comes from their ultra-low current noise. The current noise contribution of Si-JFETs is often considered negligible, however Fig.~\ref{EnIn} clearly shows that that the noise contribution is comparable to the white noise for input impedances above 10 M$\Omega$. 
The EDELWEISS collaboration has shown that current noise is dominating the ionization baseline resolution \cite{EDWreadout} of its germanium cryogenic detectors even at an optimal cold electronic working temperature of 120K.
The 30 pF HEMT current noise is about 10 times lower than the Si-JFET of similar input capacitance.

\section{Detector + HEMT Noise Model}

The previous section provided the voltage and current noise of the five studied HEMT geometries for frequencies up to 1 MHz. Additional noise sources contributions can be easily calculated with respect  to the HEMT input as done in \cite{EDWreadout} for Si-JFET. The bias circuit, detector, feedback, cabling and parasitic impedances must be included to calculate the expected measurement resolution.

Added noise from the feedback resistor can be eliminated with the use of an active reset on the HEMT gate. Added noise from the detector bias resistor can be eliminated through a similar HEMT-based active reset or by using high-value resistors located on the 20 mK detector holder. With the complete noise  and circuit model, the optimal filter formalism (details in \cite{BerkeleyLTD16}) was used to calculate the RMS baseline resolution.

\begin{figure} %[htbp]
\begin{center}
\includegraphics[%
  width=0.9\linewidth,
  keepaspectratio]{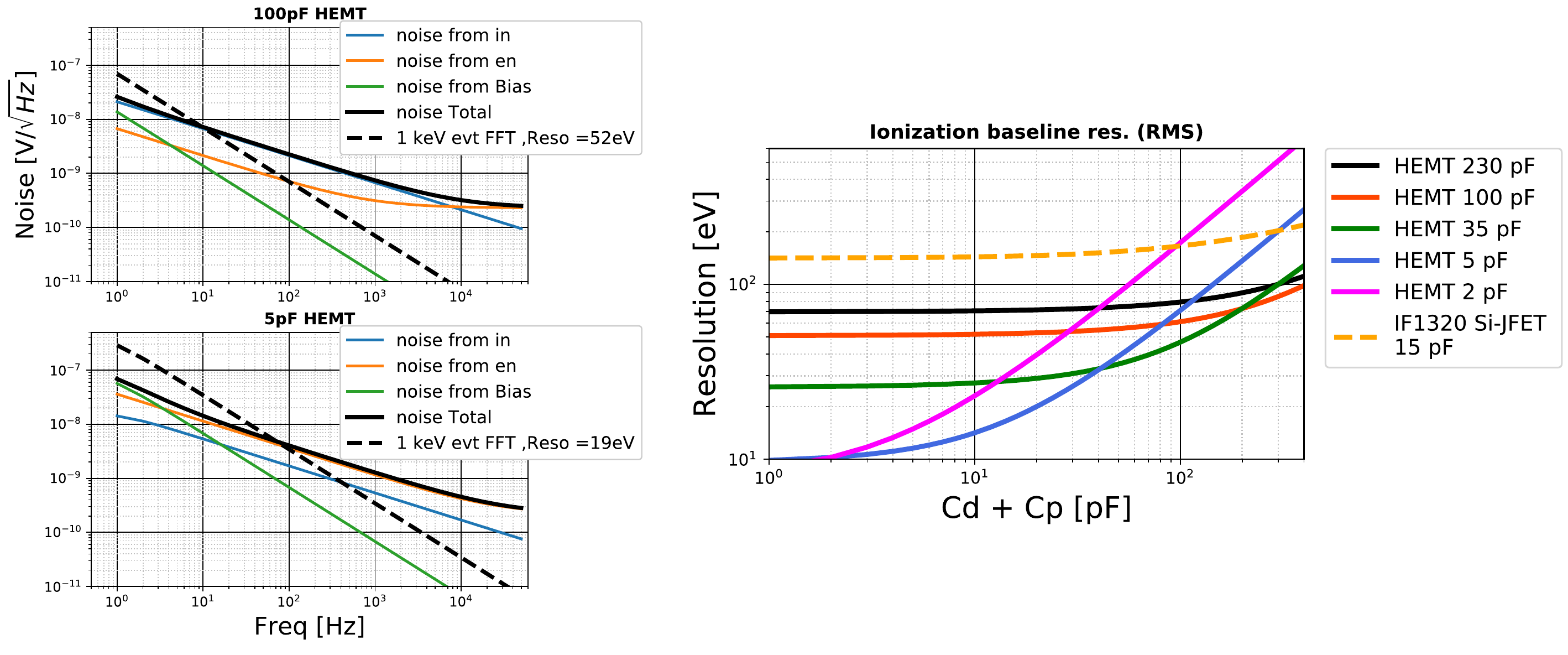}
\end{center}
\caption{{\it Left:} Contribution to the ionization resolution of the voltage noise, current noise, bias resistor (10 G$\Omega$ @ 20mK). The total detector + cabling capacitance is  20 pF. 5 pF and 100 pF geometries have been studied. FFT of 1 keV event are shown. {\it Right:} Evolution of the resolution with the detector + cabling capacitance for the 5 HEMT geometries and the IF1320 Si-JFET from InterFET. Noise of the bias resistor is included. (Color figure online.)}
\label{ResoVsCapa}
\end{figure}

Fig.~\ref{ResoVsCapa} shows the contributions to the ionization resolution of the HEMT voltage noise, current noises, thermal noise of a 10 G$\Omega$ bias resistor cooled at 20mK for a total detector and cabling capacitance of 20 pF. Results are given for two HEMT gate geometries (5 pF and 100 pF).
The current noise dominates the resolution for the 100 pF HEMT while is negligible for the 5 pF HEMT. 
Our model predicts an remarkable 19 eV$_{ee}$ rms ionization baseline resolution for the 5 pF HEMT which is one order of magnitude better than the best resolution achieved on ionisation with cryogenic detectors \cite{EDW_JINST, BerkeleyNIM}.

The same Fig.~\ref{ResoVsCapa} ({\it Right}) compiles our model predicted resolutions for the five HEMT geometries studied for the total detector + cabling input capacitance varying up to 350 pF. The thermal noise contribution from a 10 G$\Omega$ bias resistor cooled at 20 mK is included. 
Our model predicts that ionization baseline resolution below 20 eV$_{ee}$ are reachable for total a input capacitance below 20 pF with the 5 pF HEMT.

Our model shows that the basic "matching rule", according to which one should match the transistor input capacitance to the detector+cabling capacitance, is clearly insufficient for a proper optimization. As an example, with 100 pF of input load capacitance the best resolution is obtained for a 35 pF HEMT. The 100 pF HEMT performs as poorly as the 5 pF HEMT. This demonstrates the importance of the current noise, which is not used in derivation of the matching rule. A complete model of the input electronics , including both voltage and current noise, is mandatory for high impedance readout optimization.

Proper design of the cold amplifier is required to ultimately be limited by only the input HEMT noise. Previous work by some of the authors \cite{BerkeleyNIM} has demonstrated 91 eV$_{ee}$ baseline RMS ionization resolution with a HEMT-based cryogenic charge amplifier with a 240g CDMS-II cryogenic germanium detector, nearly matching the model predictions. However, the complexity required to implement a HEMT-based fully cryogenic charge amplifier for the hundreds of ionization channels present in a reasonably-sized experiment requires us to explore amplifier topologies where the use of cryogenic HEMTs is limited to only the input transistor. 

Two schemes are thus being investigating : 1) a simple follower HEMT-based stage at 1-4 K, very close to the detector with a very low noise 300 K stage and 2) an hybrid HEMT + SiGe $/$ MOS ASICs high gain stage at 1-4K. The follower scheme is used by the EDELWEISS collaboration \cite{EDW_JINST, EDWreadout} and works are in progress to adapt it to an HEMT input. A production of ASICs based on the 0.35 $\mu$m SiGe biCMOS technology from AMS foundry has been delivered and will be tested soon. This technology has demonstrated T = 4.2 K  performances \cite{ASIC_NIM}. SiGe bipolar transistors will be used for amplification and pMOS/nMOS switches will be used for calibration and reset purposes.

 \section{Conclusion : prospect for a 10 eV heat, 20 eV$_{ee}$ ionization, 35 g germanium cryogenic detector}

\begin{figure} %[htbp]
\begin{center}
\includegraphics[%
  width=0.59\linewidth,
  keepaspectratio]{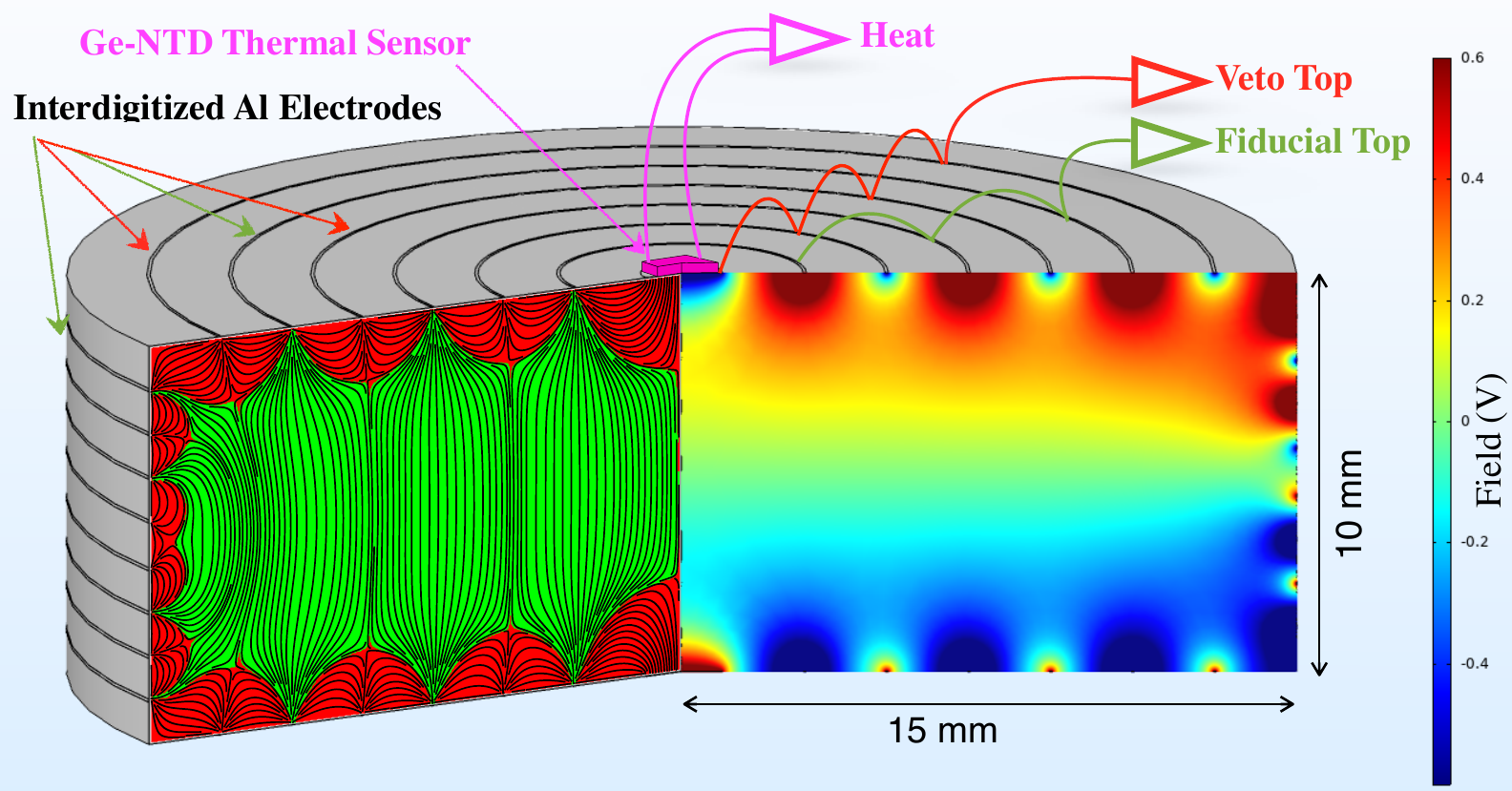}
\end{center}
\caption{Electrostatic simulation of a Full Inter-Digitized electrodes scheme on a 38 g germanium crystal ($\Phi$ = 30 g, h = 10mm). The crystal is surrounded at 2 mm distance by a chassis connected to the ground (not shown). The capacitance of the 4 electrodes with respect to the ground is about 20 pF (Color figure online.)}
\label{ComsolSimu}
\end{figure} 

The EDELWEISS collaboration has recently demonstrated an impressive 18 eV RMS heat resolution on a 34g germanium detector \cite{EDW_Surf}. Careful analysis has shown that the resolution is limited by the current noise of the Si-JFET. Our HEMT model predicts than 10 eV heat resolution would be achievable by simply replacing the JFET with a 230 pF CNRS/C2N HEMT.

Our data-driven noise model demonstrates the ability of the CNRS/C2N HEMTs to achieve extremely low energy thresholds when coupled to the appropriate detector. Electrostatic simulations (Fig. 6) show that the EDELWEISS Fully Inter-Digitized (FID) 800g detector \cite{EDW_JINST} ionization electrode pattern can be adapted to match the 20 pF capacitance requirement with a $\sim$30 g detector. This FID detector geometry has been shown to efficiently reject surface events, which may be advantageous for CE$\nu$NS and low-mass WIMP searches where the low energy backgrounds are unknown. Future work will address the issues of very low capacitance cabling and optimization of the balance between individual detector mass, fidicual volume, and detector capacitance when considering the 10 eV heat and 20 eVee ionization baseline RMS resolutions required to the upcoming Ricochet and EDELWEISS-SubGeV experiments. The development of HEMT-based amplifiers represents a clear path to achieving the energy resolution requirements of cryogenic semiconductor detectors for pursuing new physics in the low mass Dark Matter and CE$\nu$NS sectors. 

\begin{acknowledgements}
This work was supported by the LabEx Lyon Institute of Origins (ANR-10-LABX-0066) of the Universite de Lyon in the framework "Investissements d'Avenir" (ANR-11-IDEX-00007), by the Cross-Disciplinary Program on Instrumentation and Detection of CEA in the framework of the BASKET project and has received funding from the European Research Council (ERC) under the European Union's Horizon 2020 research and innovation program under Grant Agreement ERC-StG-CENNS 803079.
We thank EDELWEISS and Ricochet collaborations for their help and useful discussion in performing this work.
\end{acknowledgements}

\pagebreak

\end{document}